\documentclass[preprint,aps,pra,superscriptaddress]{revtex4-1}
\usepackage{array}
\usepackage{relsize}
\usepackage{placeins}
\usepackage{amsmath}
\usepackage{mathrsfs}
\usepackage{amssymb}
\usepackage{graphicx}
 \usepackage{color}
  \usepackage{hyperref}
\usepackage{subfigure}
\usepackage[normalem]{ulem}

\newcommand{\kMI}{{\mathcal M}^{I}}
\newcommand{\kMII}{{\mathcal M}^{II}}

\newcommand{\be}{\begin{equation}}
\newcommand{\ee}{\end{equation}} 
\newcommand{\lb}{\label}
\newcommand{\OL}{\overline}
\newcommand{\wh}{\widehat}
\newcommand{\mE}{\mathcal{E}}

\newcommand{\const}{({\rm const.})}

\newcommand{\bk}{{\bf k}}

\newcommand{\br}{{\bf r}}
\newcommand{\bs}{{\bf s}}
\newcommand{\bu}{{\bf u}}
\newcommand{\bw}{{\bf w}}

\newcommand{\bt}{{\bf t}}
\newcommand{\bv}{{\bf v}}
\newcommand{\bx}{{\bf x}}

\newcommand{\cR}{{\mathbb R}}

\newcommand{\grad}{{\mbox{\boldmath $\nabla$}}}
\newcommand{\bdot}{{\mbox{\boldmath $\cdot$}}}

\begin{document}

\title{\textbf {Extracting the Spectrum by Spatial Filtering}}
\author{Mahmoud Sadek}
\email{msadek@ur.rochester.edu}
\affiliation{Department of Mechanical Engineering, University of Rochester}
\affiliation{Irrigation and Hydraulics Department, Faculty of Engineering, Cairo University}
\author{Hussein Aluie}
\affiliation{Department of Mechanical Engineering, University of Rochester}
\affiliation{Laboratory for Laser Energetics, University of Rochester}

\begin{abstract}
We show that the spectrum of a flow field can be extracted within a \emph{local} region by straightforward filtering in physical space. We find that for a flow with a certain level of regularity, the filtering kernel must have a sufficient number of vanishing moments for the ``filtering spectrum'' to be meaningful. Our derivation follows a similar analysis by \cite{Perrieretal95} for the wavelet spectrum, where we show that the filtering kernel has to have at least $p$ vanishing moments in order to correctly extract a spectrum $k^{-\alpha}$ with $\alpha < p+2$. For example, any flow with a spectrum shallower than $k^{-3}$ can be extracted by  a straightforward average on grid-cells of a stencil. We construct two new ``simple stencil'' kernels, $\kMI$ and $\kMII$, with only two and three fixed stencil weight coefficients, respectively, and that have sufficient vanishing moments to allow for extracting spectra steeper than $k^{-3}$. We demonstrate our results using synthetic fields, 2D turbulence from a Direct Numerical Simulation, and 3D turbulence from the JHU Database. 
Our method guarantees energy conservation and can extract spectra of non-quadratic quantities self-consistently, such as kinetic energy in variable density flows, which the wavelet spectrum cannot. The method can be useful in both simulations and experiments when a straightforward Fourier analysis is not justified, such as within coherent flow structures covering non-rectangular regions, in multiphase flows, or in geophysical flows on Earth's curved surface.
\end{abstract}

\maketitle

%%%%%%%%%%%%%%%%%%%%%%%%%%%%%%%%%%%%%%%%%%%%%%%%%%%%%%%%%%%%%%%%%%%%%
% MAIN BODY OF PAPER
%%%%%%%%%%%%%%%%%%%%%%%%%%%%%%%%%%%%%%%%%%%%%%%%%%%%%%%%%%%%%%%%%%%%%
%

%% In all cases, if there is only one entry of this type within
%% the higher level heading, use the star form: 
%%
% \section{Section title}
% \subsection*{subsection}
% text...
% \section{Section title}

%vs
 \section{Introduction}
Measuring the spectrum of a flow quantifies the energy content of different spatial scales. In turbulent flows, it can yield valuable information about the cascade ranges, dissipation, and turbulence intensity. Determining the  power-law slope of a turbulence spectrum has allowed for comparison between observations and theoretical predictions \cite{Frisch95,Pope00,Davidson13}. The spectrum also conveys information about the regularity (or smoothness) of a flow field (e.g. \cite{Eyink05}).
One measure of the spectral content is the second order structure function, often employed in turbulence \cite{Anselmetetal84,Chenetal97}. It can be noisy without averaging over sufficient data points. However, its main drawback is its inability to yield the correct scaling if the field is too smooth, corresponding to a power-law spectrum steeper than $k^{-3}$ as a function of wavenumber $k$.
By far, the most common method to measure the spectrum is using Fourier transforms. However, Fourier transforms in non-periodic domains require tapering (windowed Fourier transforms) or reflecting the data to make it {\it de facto} periodic. Such alterations of the data can introduce uncontrolled errors which may compromise the results in unknown ways. For example, the usage of a tapering window introduces an artificial length-scale ---that of the window, and artificial gradients associated with the tapering which can alter the energy content of different spatial scales. Moreover, Fourier techniques are inherently global in space and cannot characterize the flow properties locally.

These shortcomings partially motivated the introduction of wavelet analysis to turbulence \citep{meneveau1991dual,meneveau1991analysis,Farge92,farge1999turbulence}. 
A main advantage wavelets have over Fourier analysis is the identification of flow properties simultaneously as a function of scale and space. As we shall elaborate below, the results presented here build upon an important work in wavelet analysis by \cite{Perrieretal95}, who established the relation between Fourier and wavelet spectra, and identified the conditions under which a wavelet spectrum is meaningful. 

Filtering (or coarse-graining) is a widely used framework in fluid dynamics, especially within the Large Eddy Simulation (LES) literature \cite{Leonard74,MeneveauKatz00,Germano92}.
It also offers a natural and versatile framework to understand the multiscale physics of complex flows (e.g. \cite{Piomellietal91,Meneveau94,Eyink95prl,Eyink05,Aluie17}). More extensive discussions of the framework and its utility can be found in many references (e.g. \cite{Eyink95,MeneveauKatz00,Chenetal03,AluieEyink10,Riveraetal14,FangOuellette16}). 
The primary goal of this paper is three-fold: (1) demonstrate that (low-pass) filtering in physical space can be used to extract the spectrum of a flow, (2) establish the relation between Fourier and filtering spectra, and (3) identify the conditions under which such a ``filtering spectrum'' is meaningful. We will also attempt to shed light on the close connections and important differences between the method employed here and wavelet analysis. 

Beyond its primary aim, we believe this work is potentially important in several respects.
It offers a new way to measure spectra that is consistent with the filtering framework used in LES and also with coarse-graining analysis of multi-scale dynamics \cite{Eyink05,Aluie13}. An important advantage of the method presented here over both Fourier and wavelet spectra is that it allows for a natural generalization of the notion of `spectrum' to non-quadratic quantities, such as kinetic energy (KE) in variable density flows, $\rho |\bu|^2/2$, as we elaborate in section \ref{sec:VDspectrum} below. It is advantageous over Fourier analysis in its ability to measure the spectrum locally in space, similar to wavelet analysis, albeit the method here is arguably simpler. This allows us, for example, to extract a spectrum within coherent flow structures covering irregularly shaped regions.  
This paper also introduces in section \ref{sec:NewKernels} generalized versions of the Top-hat kernel that are compact in physical space, are straightforward to implement on a discrete grid, and can extract spectra with a wider range of scaling exponents compared to commonly used filter kernels.

Applications that can benefit from this method include (i) 2D or 3D laboratory flow field data (e.g. from particle image velocimetry) within a window of interest,
(ii) geophysical flows on Earth's spherical surface derived from satellite measurements or from general circulation models (GCMs) \cite{Aluieetal18}. Traditional Fourier analysis on regional oceanic `boxes' is further complicated by Earth's curvature since such `boxes'  cannot be rectangular on a curved surface.
(iii) Flows from the Johns Hopkins University turbulence database (JHTDB) \cite{Lietal08}, where a user is querying a local region (as opposed to downloading the entire dataset) and can extract the spectrum using built-in filters made available through the JHTDB platform or higher-order compact kernels, such as those discussed below.
(iv) Inhomogeneous flows, such as channel flows, for which a spatially local measurement of the spectrum is desired and Fourier analysis cannot be employed without altering the data. (v) Multi-phase flows where it may be insightful to measure the spectrum conditioned on one of the phases.

The outline of this paper is as follows. In section \ref{sec:Preliminaries}, we introduce notation and preliminary material. Section \ref{sec:ExtractingSpectrum}  presents the main results, and section \ref{sec:ConnectionWavelets} discusses similarities and differences with the wavelet spectrum. Section \ref{sec:NewKernels} discusses the formulation of new filtering kernels. In section \ref{sec:ApplicationFlows}, we apply the method to realistic flows, before we conclude with section \ref{sec:Conclusion} followed by an Appendix.

\section{Preliminaries} \label{sec:Preliminaries}
In a periodic domain $\bx\in [-L/2,L/2)^d$ in $d-$dimensions, the Fourier transform and its inverse are, respectively
\begin{eqnarray}
\hat{f}(\bk) &=& \frac{1}{L^d}\int_{-L/2}^{L/2} d^d\bx~f(\bx) \,e^{-i\frac{2\pi}{L} \bk\bdot\bx}\\
f(\bx) &=& \sum_\bk  \hat{f}(\bk) \, e^{i\frac{2\pi}{L} \bk\bdot\bx}
\end{eqnarray}
This normalization guarantees that $\hat{f}(\bk=0)$ equals the spatial average, $\langle f(\bx)\rangle = {L^{-d}}\int_{-L/2}^{L/2} d^d\bx~f(\bx) \,$. 
We define the \emph{Fourier spectrum} of $f(\bx)$ as
\be E(k) = \sum_{k-\frac{1}{2}<|\bk|\le k+\frac{1}{2}} \frac{1}{2}|\hat{f}(\bk)|^2, \hspace{1cm}k = 0,1,2, \dots,
\ee
where $|\bk|$ is the Euclidean norm, $\sqrt{k_x^2+k_y^2+k_z^2}$.
The Fourier coefficients satisfy Plancherel's relation:
\be \langle \frac{1}{2}|f(\bx)|^2\rangle =\sum_{\bk}~\frac{1}{2}|\hat{f}(\bk)|^2 = \sum_{k=0}^{\infty} E(k)
\lb{eq:Plancherel}\ee

\subsection{Coarse-graining or filtering} \label{sec:coarsegraining}
For any field $u(\bx)$, a coarse-grained or (low-pass) filtered field, which contains modes
at scales $>\ell$, is defined in $d$-dimensions as
\be
\OL u_\ell(\bx) = \int d^{d}\br~ G_\ell(\bx-\br) \,u(\br),
\lb{eq:filtering}\ee
where $G(\br)$ is a normalized convolution kernel, $\int d^{d}\bs ~G(\bs)=1$, for dimensionless $\bs$. 
The kernel can be any real-valued function which decays sufficiently rapidly for large $r$. 
Its dilation in an $d$-dimensional domain, $G_\ell(\br)\equiv \ell^{-d} G(\br/\ell)$, will share these properties 
except that its main support will be in a region of size $\ell$. 

Operation (\ref{eq:filtering}) may be interpreted as a local space average in a region of size $\ell$ centered at point $\bx$.
It is, therefore, a scale decomposition performed in x-space that partitions length scales in the system into large ($\gtrsim\ell$), captured by $\OL{u}_\ell$, and small ($\lesssim\ell$), captured by the residual $u'_\ell=u-\OL u_\ell$. This is perhaps made clearer by considering the filtered field in k-space,
\begin{eqnarray} 
\OL{u}_\ell(x) = \int^{+\infty}_{-\infty} dk~ \wh{G}(k\ell) \,\wh{u}(k) \, e^{ikx}, 
\lb{eq:FilteredFieldKspace}
\end{eqnarray} 
where 
\begin{eqnarray}
\wh{G}(\ell k)= \wh{G_\ell}(k) = \frac{1}{L}\int_{-L/2}^{L/2} dx~G_\ell(x) \,e^{-i\frac{2\pi}{L} kx}~~.\nonumber
\end{eqnarray}
From eq. (\ref{eq:FilteredFieldKspace}), the spectrum of the filtered field, $\OL{u}_\ell$, is
\be
\tilde{E}(k) = |\wh{G}(k\,\ell)|^2~ E(k),
\ee
such that $\tilde{E}(k)$ and $E(k)$ agree at low wavenumbers since $\wh{G}(k)\to1$ as $k\to0$. $\tilde{E}(k)$ decays rapidly beyond the cutoff waveneumber, $k_\ell$, associated with the filtering scale $\ell$; {\it i.e.} for $k \gtrsim k_\ell=L/\ell$, where $L$ is the domain size. In the limit of small filtering length, $\ell \to 0$, the spectrum of the filtered field approaches that of the unfiltered spectrum, $\tilde{E}(k) \to E(k)$, as would be expected. 

Therefore, it seems it should be possible to infer the spectrum of a field by filtering at different length-scales, 
\emph{without} performing Fourier transforms.

\section{Extracting the spectrum \lb{sec:ExtractingSpectrum}}
By varying the filter scale $\ell$ and measuring the corresponding change in the cumulative energy, $|\OL{\bu}_\ell|^2/2$, we will show that it is indeed possible to infer the spectral content at scale $\ell$. In fact, implementing this procedure with a sharp-spectral filter kernel yields exactly the Fourier spectrum of a signal as discussed in \cite{Frisch95}. However, filtering with a sharp spectral filter requires performing Fourier transforms, whereas the goal here is extracting the spectrum while in x-space.  We will use the term `filtering spectrum' to refer to the spectrum obtained by filtering in x-space and to distinguish it from the `Fourier spectrum' obtained by traditional Fourier transforms in k-space. 

To justify comparing the slopes of filtering spectra to those from turbulence theory or to global Fourier spectra, it is necessary to ensure that  slopes are consistent. As we shall elaborate in this section, the kernel $G(r)$ has to satisfy certain conditions to guarantee that the filtering spectrum conveys the correct energy content at different length-scales.

\subsection{Kernel properties}
We shall assume that the kernel is a real-valued even function, $G(\br) = G(-\br)$. Hence, its Fourier transform, $\wh{G}(\bk)$, will also be real-valued. Any spherically symmetric kernel is even. We also assume that the filter kernel is normalized, $\int d\br \,G(\br) = \wh{G}(\bk=0) = 1$.

In practical applications, filtering kernels are often chosen to be sufficiently localized in x-space to avoid prohibitive computational costs. Therefore, we shall restrict our consideration to kernels that decay faster than any power in x-space,
$G(\br) \le \const r^{-m}$ for any $m$ as $|\br| \to \infty$, where $r = |\br| = \sqrt{x^2+y^2+z^2}$ is the Euclidean norm. Examples of such kernels include the Gaussian, $(\frac{1}{2\pi})^{3\over2}e^{-|\br|^2/2}$, or any kernel that has compact support (i.e. has zero value beyond a finite spatial extent) such as the Top-hat kernel,
\begin{eqnarray}
H_\ell(x)&=&\begin{cases}
    1/\ell, & \text{if $|x|<\ell/2$}.\\
    0, & \text{otherwise}.\\
  \end{cases}\lb{app_eq:Tophat_x}
\end{eqnarray}
These kernels are well-localized in x-space which makes them useful in practical problems. They are also useful analytically since the fast decay in x-space guarantees smoothness in k-space. A Taylor-series expansion near the origin in k-space yields:
\be
\wh{G}(k) = \wh{G}(0) + k\,\wh{G}^{(1)}(0) + k^2\,\frac{\wh{G}^{(2)}(0)}{2!} + \dots
\lb{eq:TaylorExpandKernel_1}\ee
where $f^{(n)}(s)$ denotes the $n$-th derivative, $\frac{\partial^n}{\partial s^n} f(s)$.

Moments of a kernel are related to its derivatives in k-space:
\be \int^{+\infty}_{-\infty} dx ~x^n \,G(x) = \wh{G}^{(n)}(k)\Big|_{k=0}.
\lb{eq:MomentsDerivatives}\ee
Since even kernels, $G(x)=G(-x)$, have vanishing odd moments, it follows from eq. (\ref{eq:MomentsDerivatives}) that 
 $ \wh{G}^{(n)}(0)=0$ for all odd integers $n$ for any even kernel $G(x)$. 

We shall call a kernel $G(x)$ ``$p$-th order'' iff
\begin{eqnarray} 
 \int^{+\infty}_{-\infty}  dx ~x^n G(x) &=& 0 \hspace{1cm} \mbox{for} \hspace{1cm} n=1,\dots,p, \nonumber\\
 \mbox{and}\hspace{1cm} \int^{+\infty}_{-\infty}  dx ~x^{p+1} G(x) &\ne& 0.
\lb{eq:pthOrderKernel}
\end{eqnarray} 
Any even kernel is of an odd integer order $p\ge 1$. For example, the Gaussian and Top-hat kernels are of order $p=1$. As we will discuss below, the order of the kernel is a key property for extracting the correct spectrum by filtering. We will also show how to construct simple kernels of any order.

For a normalized even $p$-th order kernel, the Taylor expansion in eq. (\ref{eq:TaylorExpandKernel_1}) becomes
\begin{eqnarray} 
\wh{G}(k) &=& 1 + k^{p+1}\underbrace{\left[\frac{\wh{G}^{(p+1)}(0)}{(p+1)!}  + k^{2}\,\frac{\wh{G}^{(p+3)}(0)}{(p+3)!}  + \dots \right]}_{\phi(k)}, \lb{eq:TaylorExpandKernel}
\end{eqnarray} 
where
\begin{eqnarray} 
\phi(0) = \const\ne 0. 
\lb{eq:PhiProperty0}\end{eqnarray}

Note that in the Taylor expansion in eq. (\ref{eq:TaylorExpandKernel}), we are using smoothness properties (in k-space) of the kernel and not of the field being filtered. 

\subsection{The Filtering Spectrum}
We define the \emph{filtering spectrum} as
\be
\OL{E}(k_\ell) \equiv \frac{d}{dk_\ell} \left\langle |\OL{u}_\ell(\bx)|^2\right\rangle/2 = -\frac{\ell^2}{L}\frac{d}{d\ell}\left\langle |\OL{u}_\ell(\bx)|^2\right\rangle/2,
\lb{eq:DefFilteringSpectrum}\ee
where $k_\ell=L/\ell$ over a domain or region of interest with characteristic length $L$. It agrees with the traditional Fourier spectrum under certain conditions, as we shall show. 
We also define the \emph{cumulative spectrum}, following \cite{Frisch95}, as
\be
\mE(k_\ell) \equiv \frac{1}{2}\left\langle |\OL{u}_\ell(\bx)|^2\right\rangle.
\lb{eq:DefCumulativeSpectrum}\ee

We work in 1-dimension to simplify the presentation. The filtering spectrum in eq. (\ref{eq:DefFilteringSpectrum}) can be expressed as
\begin{eqnarray}
\OL{E}(k_\ell) = \int_0^{\infty} dk~ \left[\frac{d}{dk_\ell}  \left|\wh{G}\left(\frac{k}{k_\ell}\right)\right|^2\right] E(k).
\lb{eq:FilteringFourierSpectrum_Rel2}\end{eqnarray}
Note that the function multiplying the Fourier spectrum, $E(k)$, in the integrand,
\be
F(k) \equiv  \frac{d}{dk_\ell}  \left|\wh{G}\left(\frac{k}{k_\ell}\right)\right|^2,
\nonumber\ee
gets dilated (becomes broader) as a function of $k$ when $k_\ell \to \infty$. For any fixed $k_\ell$, the function $F(k)$ is a hump that is approximately localized around $k\sim k_\ell$ and tends to zero for $k\ll k_\ell$ and $k\gg k_\ell$. This leads to the averaging integral in eq. (\ref{eq:FilteringFourierSpectrum_Rel2}) deriving most of its contribution from wavenumbers $k$ in the vicinity of $k_\ell$.
However, the larger is $k_\ell$, the broader is the averaging integral due to the broadening of $F(k)$. 
This is analogous to what happens in a wavelet spectrum, as highlighted by Perrier et al. \cite{Perrieretal95}, who observed that the 
scaling of the wavelet (or filtering, in our case) spectrum at large $k_\ell$ depends on the properties of the wavelet (or filtering kernel, in our case) 
at small wavenumbers.

Following the analysis of Perrier et al. \cite{Perrieretal95} for the wavelet spectrum, 
assume that $E(k)= \const k^{-\alpha}$ over $k_a<k<\infty$, then the filtering spectrum using a $p$-th order kernel scales as
\begin{eqnarray}
\OL{E}(k_\ell)&=&\int_0^{\infty} dk~ \frac{d}{dk_\ell}  \left|\wh{G}\left(\frac{k}{k_\ell}\right)\right|^2 E(k) \nonumber\\
&=&\underbrace{ \int_0^{k_a} dk~ \frac{d}{dk_\ell}  \left|\wh{G}\left(\frac{k}{k_\ell}\right)\right|^2 E(k)}_{\hspace{1cm}\mbox{\footnotesize{term 1}}~\sim k_\ell^{-(p+2)}} + \underbrace{\const \int_{k_a}^{\infty} dk~ \frac{d}{dk_\ell}  \left|\wh{G}\left(\frac{k}{k_\ell}\right)\right|^2 k^{-\alpha} }_{\hspace{1cm}\mbox{\footnotesize{term 2}}~\sim k_\ell^{-\alpha}}
\lb{eq:FilterSpectrumScaling_1}\end{eqnarray}
The derivation is provided in the Appendix below.
Eq. (\ref{eq:FilterSpectrumScaling_1}) implies that if the Fourier spectrum decays faster than $k^{-(p+2)}$, the small-wavenumber contributions in `term 1' dominate at large $k_\ell$, whereas if $\alpha < p+2$, then the `filtering spectrum' has the same power-law slope as the Fourier spectrum.

We conclude that if the Fourier spectrum has a power-law scaling $E(k)\sim k^{-\alpha}$ at high wavenumbers, then the `filtering spectrum' obtained by filtering with a $p$-th order kernel $G(r)$ scales as 
\begin{eqnarray}
\OL{E}(k)&\sim&\begin{cases}
    k^{-\alpha}, & \text{if $\alpha < p+2$}\\
    k^{-(p+2)}, & \text{if $\alpha> p+2$}\\
  \end{cases}\lb{eq:FilterSpectrumScaling}
\end{eqnarray}
Therefore, the steeper is the underlying spectrum, the higher is the order of the filtering kernel required for extracting such a spectrum.
For example, the Gaussian or Top-hat functions are 1st-order kernels and can only extract power-law spectra shallower than $k^{-3}$. As Fig. \ref{fig:GaussTopSpectra} below shows, if the Fourier spectrum decays faster than $k^{-3}$, the filtering spectrum will `lock' at a $k^{-(p+2)}$ scaling for kernels with $p=1$. A practical consequence of eq. (\ref{eq:FilterSpectrumScaling}) is that if a filtering spectrum is measured using a $p$-th order kernel and exhibits a scaling shallower than $k^{-(p+2)}$, then the user can have confidence that it reflects the scaling of the Fourier spectrum correctly. Otherwise, if it scales $\sim k^{-(p+2)}$, then a higher order filtering kernel is required. It should be noted that the scaling of the filtering spectrum derived in relation (\ref{eq:FilterSpectrumScaling}) is asymptotic. The agreement of $\OL{E}(k)$ with the Fourier spectrum may not be perfect in practice, even when using a kernel satisfying $\alpha < p+2$. This is due to a limited range of scales and the fact that compact spatial kernels are not strictly local in k-space as the sharp-spectral filter, which can lead to additional smoothing as a function of scale (e.g. \cite{Riveraetal14,McCaffreyetal15}).

 \begin{figure}[bhp]
{\includegraphics[width=0.44\textwidth,height=0.2\textheight]{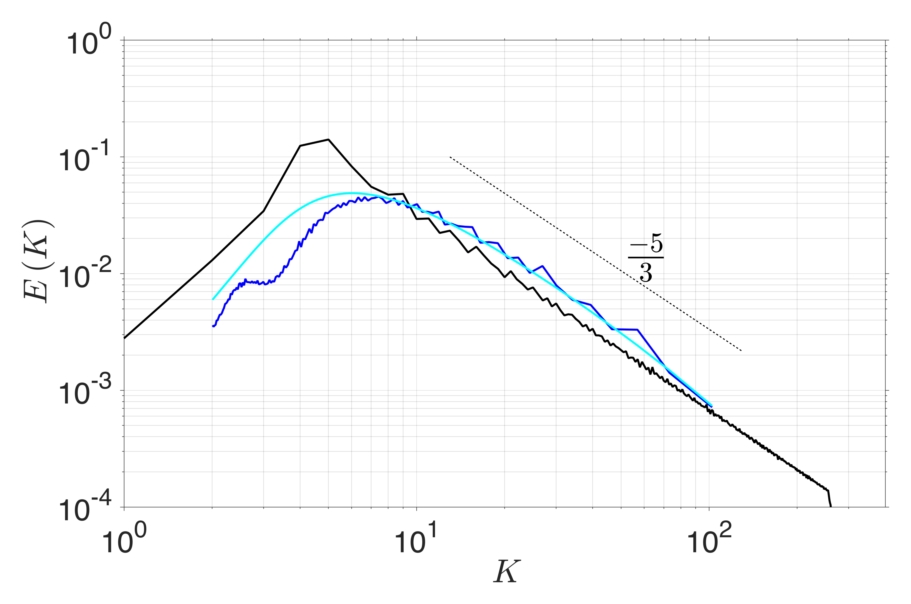} 
\includegraphics[width=0.44\textwidth,height=0.2\textheight]{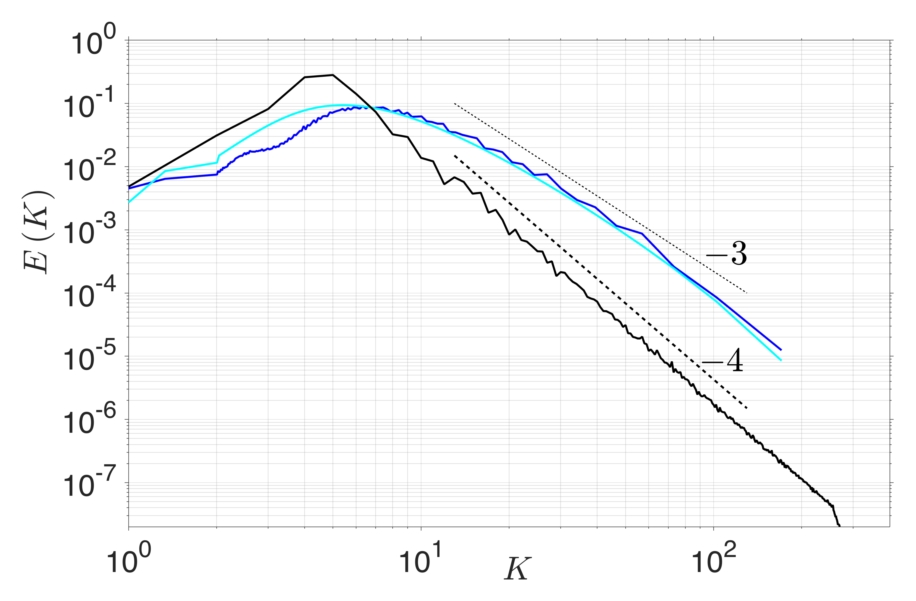}
\includegraphics[width=0.44\textwidth,height=0.2\textheight]{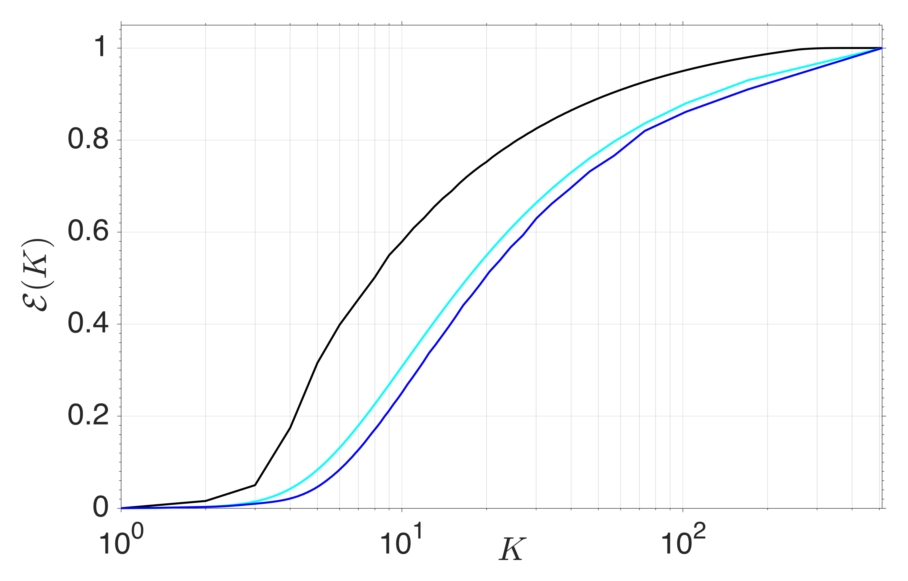}
\includegraphics[width=0.44\textwidth,height=0.2\textheight]{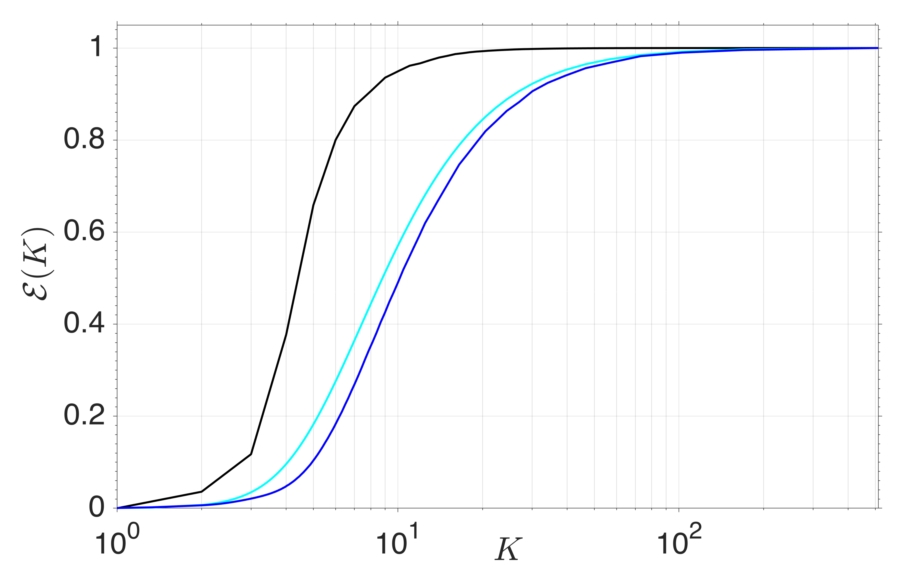}}
\caption{\footnotesize{Comparing Fourier and filtering spectra for a synthetic doubly-periodic velocity field  with a preset Fourier power-law spectrum (black): $E(K)\sim K^{-5/3}$ (upper left panel) and $E(K)\sim K^{-4}$ (upper right panel). The filtering spectra, $\OL{E}\left(K\right)$ are computed by convolving the velocity in x-space with Top-hat (dark blue) and Gaussian (light blue) kernels. These two kernels correctly extract the $k^{-5/3}$ scaling, but not the $k^{-4}$, locking-in at a $k^{-3}$ scaling instead as described by eq. (\ref{eq:FilterSpectrumScaling}).
Lower two panels show the corresponding cumulative spectra, $\mE(K)$, from which the filtering spectra are derived (eqs. (\ref{eq:DefFilteringSpectrum})-(\ref{eq:DefCumulativeSpectrum})). Note that using any kernel, $\mE(K)$ converges to the total energy (normalized to unity) at large $K$, demonstrating energy conservation (eq.(\ref{eq:EnergyConservation_2})). 
}}	\label{fig:GaussTopSpectra}\end{figure}

\subsection{Energy Conservation}
It is straightforward to verify that the integral of the filtering spectrum yields the total energy:
\begin{eqnarray}
\frac{1}{2}\left\langle |u|^2 \right\rangle = \frac{1}{2}\left\langle| \OL{u}_{\ell_0}|^2 \right\rangle+\int_{k_{\ell_0}}^\infty dk_\ell~\OL{E}(k_\ell), 
\lb{eq:EnergyConservation_1}\end{eqnarray}
where $\ell_0=L/k_{\ell_0}$ is the largest filtering length-scale used. In principle, $\ell_0$ can be arbitrarily large, even larger than $L$, the size of the domain of interest. 
In practice, since the spectrum itself is calculated from the \emph{cumulative spectrum}, $\mathcal{E}(k_\ell)$, the total energy is retrieved by taking the limit of small filter scale:
\be \lim_{k_\ell\to \infty}\mE(k_\ell) = \frac{1}{2}\left\langle |u|^2 \right\rangle
\lb{eq:EnergyConservation_2}\ee
Eqs. (\ref{eq:EnergyConservation_1})-(\ref{eq:EnergyConservation_2}) show that the spectral densities obtained with filtering integrate to the total energy. This is similar to the wavelet spectrum and justifies the calculation of a spatially \emph{local} filtering spectrum to characterize and compare spectral properties at different locations in x-space. 

\subsection{Positive Definiteness}\lb{sec:PositiveDefiniteness}
One property of the filtering spectrum worth discussing is its positive semi-definiteness; from eq. (\ref{eq:FilteringFourierSpectrum_Rel2}), it is can be seen that $\OL{E}\left(k_\ell\right) \ge 0$ is guaranteed if the power spectrum of the filtering kernel itself decays monotonically in k-space, 
\be
\frac{d}{dk}|\wh{G}(k)|^2\le 0, \hspace{1cm} \mbox{for}\hspace{.5cm} k\in(0,\infty).
\lb{eq:PositiveCondition}\ee
Condition (\ref{eq:PositiveCondition}) is sufficient but not necessary to guarantee $\OL{E}\left(k_\ell\right) \ge 0$. For example, both the gaussian and sharp spectral filters satisfy this condition, but not the Top-hat kernel. 

We now offer an alternate argument on why both the Top-hat and Gaussian kernels should yield $\OL{E}\left(k_\ell\right) \ge 0$. It is known (e.g. \cite{LiebLoss01}) that if a  function $J:K\to \cR$, where $K\subseteq\cR^n$, is convex, then for any $\bv\in K$ there exists a vector $\bt\in\cR^n$ such that
\be J(\bw) - J(\bv) \ge \bt\bdot(\bw-\bv), \hspace{0.5cm}\mbox{for every}~\bw\in K.
\lb{eq:convex}\ee
The vector $\bt=\bt(\bv)$ lies in a support plane at point $\bv$. By taking, $J(\bu)=|\bu|^2$, $\bw = \OL{\bu}_\ell(\bx+\br)$, and $\bv=\OL\bu_\Delta(\bx)$, we have
\be |\OL{\bu}_\ell(\bx+\br)|^2 \ge |\OL{\bu}_\Delta(\bx)|^2 + \bt\left[\OL{\bu}_\Delta(\bx)\right]\bdot\left[\OL{\bu}_\ell(\bx+\br) - \OL{\bu}_\Delta(\bx)\right]
\lb{eq:FilterIneq_1}\ee
Consider the kernel $\Gamma(\br) = G_\ell^{-1}*G_\Delta$. It should satisfy $\Gamma(\br)\ge0$ if $G(\br)$ is concave (which implies $G\ge0$) and $\Delta\ge\ell$ as we discuss in the Appendix. We have
\be \OL\bu_\Delta(\bx) = \Gamma * \OL\bu_\ell(\bx)  = \int d^n\br \, \Gamma(\br) \,\OL\bu_\ell(\bx+\br).
\lb{eq:SoftDeconv}\ee
Operation (\ref{eq:SoftDeconv}) is a ``soft deconvolution'' (e.g. \cite{Sagaut06}), where we  deconvolve, $G_\ell^{-1}*$, at scales larger than $\Delta\ge\ell$. An interesting limit is when $\ell\to 0$, such that $\Gamma \to G_\Delta\ge0$. Another relevant limit is when $\ell\to \Delta$, yielding $\Gamma \to 1\ge0$. A more detailed discussion is provided in the Appendix.

Multiplying eq. (\ref{eq:FilterIneq_1}) by $\Gamma(\br)$ and integrating over $\br$ yields
\be  \Gamma* |\OL{\bu}_\ell|^2(\bx) \ge \int d^n\br \, \Gamma(\br) ~|\OL{\bu}_\Delta|^2(\bx) + \bt\bdot\left[\Gamma*\OL{\bu}_\ell(\bx) - \OL{\bu}_\Delta(\bx)\right] = |\OL{\bu}_\Delta|^2(\bx).
\lb{eq:FilterIneq_2}\ee
In the limit $\ell\to0$, eq. (\ref{eq:FilterIneq_2}) reduces to a positive pointwise sub-grid kinetic energy, $(\OL{|\bu|^2}_\Delta-|\OL{\bu}_\Delta|^2)/2\ge0$, as shown in \cite{Vremanetal94}.
Finally, space averaging over $\bx$ shows that the cumulative spectrum is a monotonically increasing function of $k_\ell = L/\ell$:
\be \frac{1}{2}\langle|\OL{\bu}_\ell|^2\rangle=\mE(k_\ell) \ge \mE(k_\Delta)=\frac{1}{2}\langle|\OL{\bu}_\Delta|^2\rangle, \hspace{.5cm}\mbox{for}~k_\ell\ge k_\Delta.
\lb{eq:MonotonicCumSpectrum}\ee
Therefore, we expect the filtering spectrum, being the derivative of $\mE(k_\ell)$, to be positive when using concave (and, therefore, positive) kernels.
If either (i) condition (\ref{eq:PositiveCondition}) or (ii) concavity of $G(r)$ are not satisfied by the kernel, then it is possible for $\OL{E}\left(k_\ell\right)$ to have negative values, although this does not seem to be an issue from the cases we analyze in this paper. By definition, the cumulative spectrum, $\mE(k_\ell)$, is always positive.

\subsection{Energy Spectrum in Variable Density Flows}\lb{sec:VDspectrum}
Kinetic energy in flows with variable density, such as in compressible or multiphase flows, is not quadratic but cubic due to a spatially varying density field. How can one measure its ``energy spectrum''?
Traditionally, this has been done by calculating $|\wh{\sqrt{\rho}\bu}(k)|^2$ \cite{KidaOrszag90}, which circumvents the problem by decomposing kinetic energy as if it were quadratic.  However, as demonstrated by \cite{ZhaoAluie18}, such a decomposition violates the so-called \emph{inviscid criterion}, yielding difficulties with disentangling viscous from inertial dynamics in turbulent flows. 
A Favre decomposition, $|\OL{\rho\bu}_\ell|^2/2\OL{\rho}_\ell$, respects the cubic nature of kinetic energy and satisfies the inviscid criterion, which allows for disentangling the scale dynamics clearly \cite{ZhaoAluie18}. 

The filtering spectrum presented here generalizes naturally to quantities with a nonlinearity higher than quadratic. A cumulative spectrum in variable density flows that is consistent with the Favre decomposition can be defined as
\be
\mE^F(k_\ell) \equiv \frac{1}{2} \left\langle\frac{|\OL{\rho\bu}_\ell|^2}{\OL{\rho}_\ell} \right\rangle,
\lb{eq:DefCumulativeSpectrum_Favre}\ee
and the associated filtering spectrum is
\be
\OL{E}^F(k_\ell) \equiv \frac{d}{dk_\ell} \mE^F(k_\ell).
\lb{eq:DefFilteringSpectrum_Favre}\ee

The filtering spectrum should be positive following an argument similar to that of the preceding subsection. Taking $\bw = \OL{\rho\bu}_\ell(\bx+\br)/\OL\rho_\ell(\bx+\br)$, and $\bv=\OL{\rho\bu}_\Delta(\bx)/\OL\rho_\Delta(\bx)$ in eq. (\ref{eq:convex}), we have
\be \frac{|\OL{\rho\bu}_\ell(\bx+\br)|^2}{|\OL\rho_\ell(\bx+\br)|^2} \ge \frac{|\OL{\rho\bu}_\Delta(\bx)|^2}{|\OL\rho_\Delta(\bx)|^2} + \bt\bdot\left[\frac{\OL{\rho\bu}_\ell(\bx+\br)}{\OL\rho_\ell(\bx+\br)} - \frac{\OL{\rho\bu}_\Delta(\bx)}{\OL\rho_\Delta(\bx)}\right].
\lb{eq:FavreFilterIneq_1}\ee
Multiplying eq. (\ref{eq:FavreFilterIneq_1}) by $\Gamma(\br)\,\OL\rho_\ell(\bx+\br)$, where $\Gamma(\br)= G_\ell^{-1}*G_\Delta$ is the same as in eq. (\ref{eq:SoftDeconv}), then integrating over $\br$ yields
\be \Gamma*\frac{|\OL{\rho\bu}_\ell(\bx)|^2}{\OL\rho_\ell(\bx)} \ge \frac{|\OL{\rho\bu}_\Delta(\bx)|^2}{\OL\rho_\Delta(\bx)} + 0.
\lb{eq:FavreFilterIneq_2}\ee
Note that $\OL\rho_\ell>0$ when using kernels $G\ge0$, where $\OL\rho_\ell\ne0$ in the absence of vaccum, $\rho(\bx)\ne0$ \cite{Aluie11,Aluie13}.
Finally, space averaging over $\bx$ gives that the cumulative spectrum is a monotonically increasing function of $k_\ell = L/\ell$ and, therefore, $\OL{E}^F(k_\ell)\ge0$ when using a concave filter kernel such as the Gaussian or Top-hat. Further discussion and results on this generalized energy spectrum in variable density flows will be presented in forthcoming work \cite{Zhaoetal_inprep}.

\subsection{Numerical Implementation}\lb{sec:numerics}
Here, we describe how we calculate $\OL{E}(k_\ell)$ to produce the plots presented in this paper (Figs. \ref{fig:GaussTopSpectra}, \ref{fig:NewKernelSpectrum}, \ref{fig:2Dspectrum}-\ref{fig:Strain}). We filter the velocity at every grid-point over a sub-domain of interest at two slightly different length-scales, $\ell_1$ 
and $\ell_2$. We require $\ell_1$ and $\ell_2$ to be even integer multiples of the grid-cell size $\Delta x$ to ensure that the filtering kernel is properly resolved on the grid. This is necessary to guarantee that the discretized kernel (or stencil) moments vanish to within high precision. 
More precisely,
\be \ell_m = 2\,m \,\Delta x,  \hspace{1cm}m=1,2,...
\ee
The factor $2$ ensures that even kernels yield vanishing odd moments.
Wavenumbers associated with a scale $\ell_m$ are not necessarily integers,
\be k_{\ell_m} = \frac{L}{\ell_m}. 
\lb{eq:FilteringWavenumber}\ee
This is why  plots of $\OL{E}(k)$ presented here have data points at fractional wavenumbers, for example at $k=2.3$ in Fig. \ref{fig:2Dspectrum}, whereas the Fourier spectra do not. Note that $L$ in the definition of wavenumber in eq. (\ref{eq:FilteringWavenumber}) can be replaced with any other reference scale, such as the size of a sub-domain of interest.

After filtering, the cumulative spectrum is calculated by averaging the energy 
over a sub-domain of interest, $\mE(k_\ell) = 0.5\langle|\OL{u}_\ell|^2\rangle_{\mbox{\tiny{sub-domain}}}$.
The filtering spectrum is finally obtained from a finite difference, 
\be
\OL{E}(k_{\ell_m}) = \frac{\mE(k_{\ell_{m}})- \mE(k_{\ell_{m-1}})}{k_{\ell_{m}}-k_{\ell_{m-1}}}.
\ee

\section{Connection to Wavelets\lb{sec:ConnectionWavelets}}

There is a direct connection between filtering and wavelet analysis, the latter being a band-pass filtering with the proper choice of the filtering kernel.
After all, a wavelet is constructed from a low-pass filter kernel called the scaling function. Wavelets have been used in many studies of turbulence to overcome some of the Fourier transform's drawbacks. Meneveau \cite{meneveau1991dual,meneveau1991analysis} employed orthogonal wavelets to study three-dimensional turbulence as a function of both space and length scale. Meneveau's analysis showed that kinetic energy in a turbulent flow is highly intermittent in space, exhibiting fractal scaling, and that the scale-transfer fluctuates greatly in x-space.
Wavelets have also been used extensively by Farge, Schneider, and collaborators \citep{Farge92,farge1992improved,farge1996wavelets,farge1999turbulence} 
to investigate local properties of turbulence such as energy density, spectrum, and intermittency.  
They have also been implemented in compression algorithms for turbulence research \cite{Fargeetal00}, and to isolate coherent vortex tubes from the incoherent background using orthogonal wavelets. It is beyond our scope here to review more recent studies using wavelets and similar functions, such as curvelets, (e.g. \cite{YamadaOhkitani91,KatulParlange95,VasilyevPaolucci97,GoldsteinVasilyev04,Bermejo-MorenoPullin08,Maetal09,Xiaetal09}), but instead refer the reader to more dedicated surveys \cite{Farge92,Jaffard01,SchneiderVasilyev10,FargeSchneider15,Pulidoetal16}.

The filtering spectrum presented here is in several aspects similar to that obtained with continuous and discrete wavelet transforms (CWT and DWT, respectively), but there are important differences. Non-orthogonal wavelets in the DWT they cannot be used to calculate the wavelet spectrum since they do not satisfy Plancherel's relation, {\it i.e.} energy conservation. 
Orthogonality is a significant restriction on the type of functions that can be used. For example, the Gaussian or Simple Stencil kernels we introduce below are not orthogonal. Orthogonal wavelets with compact support, such as the Haar and Daubechies wavelets, constitute a relatively small class of functions \cite{Strang89,ChuiJiang13}. The requirement of both orthogonality and compact support results in irregular functions which cannot be symmetric (except for the Haar scaling function) \cite{Daubechies88}, and which cannot be represented as explicit functions in closed form, relying instead on recursive algorithms for their evaluation \cite{Daubechies92}.

Unlike the DWT, the CWT conserves energy without requiring orthogonality of the wavelet functions. Energy conservation by the CWT relies on the Plancherel relation (\ref{eq:Plancherel}), which is limited to when energy is treated as a quadratic quantity (e.g. \cite{GrossmannMorlet84,Mallat99}). In contrast, since the filtering spectrum is the derivative of the cumulative spectrum  [eq. (\ref{eq:DefFilteringSpectrum})], it conserves energy due to the Fundamental Theorem of Calculus. This allows the filtering spectrum to analyze energy with a nonlinearity higher than quadratic as discussed in section \ref{sec:VDspectrum} above.

As we mentioned earlier, our results and the derivation in the Appendix build upon the proofs of Perrier et al (1995) who showed that in order for a wavelet spectrum to correctly capture a power-law spectrum 
$E(k)\sim k^{-\alpha}$, the analyzing wavelets are required to have at least the first $p$ moments vanish, 
such that $\alpha \le 2p+1$ (compared to a filter kernel in eq. (\ref{eq:pthOrderKernel}) above, a wavelet $\psi(x)$ is said to have $p$ vanishing moments if, $\int^\infty_{-\infty} dx \, x^{n} \psi(x) = 0$ for $n=0,\dots,p-1$, and $\int dx \, x^{p} \psi(x)\ne 0$. Note that $n=0$ is counted as a moment, unlike that for a normalized low-pass filter kernel which has to evaluate to unity). This is to be contrasted with the condition on the filtering kernels we derive here, $\alpha \le p+2$. Therefore, for a given number of vanishing moments $p$,  wavelets can correctly extract steeper spectra than is possible with the filtering spectrum. However, for even  kernels, $G(\br)=G(-\br)$, odd moments automatically vanish and $p$ increases in increments of two, eliminating the advantage of wavelets in this regard.

\section{Constructing Simple Stencil Kernels of Higher Order}\lb{sec:NewKernels}

Result (\ref{eq:FilterSpectrumScaling}) underscores the need to filter with kernels of a sufficiently high order to correctly capture the spectral scaling. 
Note that any kernel of order higher than unity cannot be positive everywhere in x-space, otherwise even moments would not vanish. We will now show how to construct simple kernels of any order.

In mathematical parlance, a `simple function' is a function that evaluates to a finite number of constants over different subsets of the domain  (e.g. \cite{LiebLoss01}). The Top-hat kernel is an elementary example of a simple function, taking on values of 1 and 0 over the domain.
Fig. \ref{fig:NewKernels} below shows two other examples of simple functions that have compact support ({\it i.e.} are zero beyond a finite extent). The advantage of using compact simple functions as filtering kernels is that their dilation, corresponding to different lengthscales $\ell$, is fairly easy to implement numerically. These kernels can be thought of as grid stencils with a finite number of predetermined weights. Simple stencil (hereafter, SS) kernels of any order can be constructed in a straightforward manner with their moments made to vanish with high accuracy.

The two SS kernel we introduce, which we label $\kMI$ and $\kMII$ and are shown in Fig. \ref{fig:NewKernels}, are 3rd and 5th order, respectively.
$\kMI$ and $\kMII$ take on two and three non-zero piecewise constant values, respectively. 
 
Let us consider how to construct $\kMI$, which we require to be symmetric, normalized, simple, compactly supported, and have a vanishing 2nd moment. The last requirement necessitates that the kernel have negative values. The most elementary kernel satisfying these conditions is one
with a main `body' having a positive value $c$, and a `leg' on either side having a negative value $-a$, as shown in Fig. \ref{fig:NewKernels}. The width of its main body is taken to be $\ell$, associated with the filtering length-scale. Parameter $c$ is set by normalization. This leaves two parameters, $a/c$ and the leg width, $b$. The former is determined from $\int dx~ x^2 \kMI(x) = 0$, which yields
\be
\hspace{2cm}\frac{a}{c}= \frac{ 1}{\left(1+2b/\ell\right)^3-1} \hspace{2cm}(\kMI~\mbox{parameter} )
\lb{eq:MI_Aconstraint}\ee
The second parameter, $b$, is free. In this paper, we take $b=\ell/8$, which yields $a/c = 64/61$ from eq. (\ref{eq:MI_Aconstraint}) and $c=\ell^{-1} 61/45$ from normalization. 

\begin{figure}[bhp]
\centering
		\includegraphics[width=1\textwidth,height=0.3\textheight]{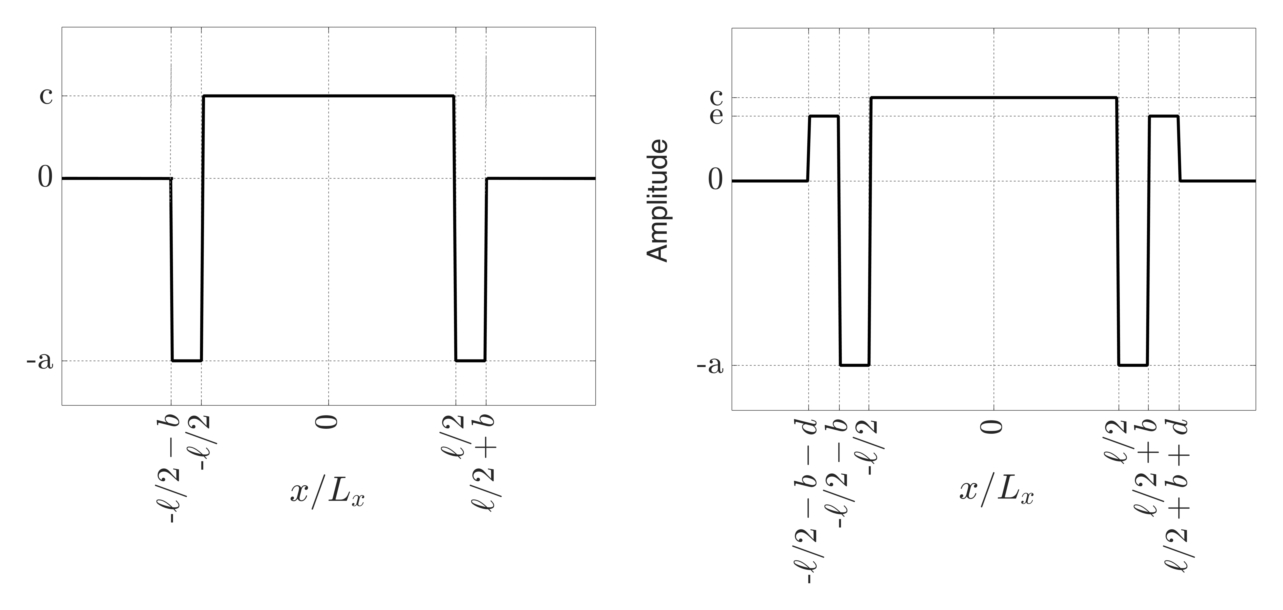}
		\includegraphics[width=1\textwidth,height=0.3\textheight]{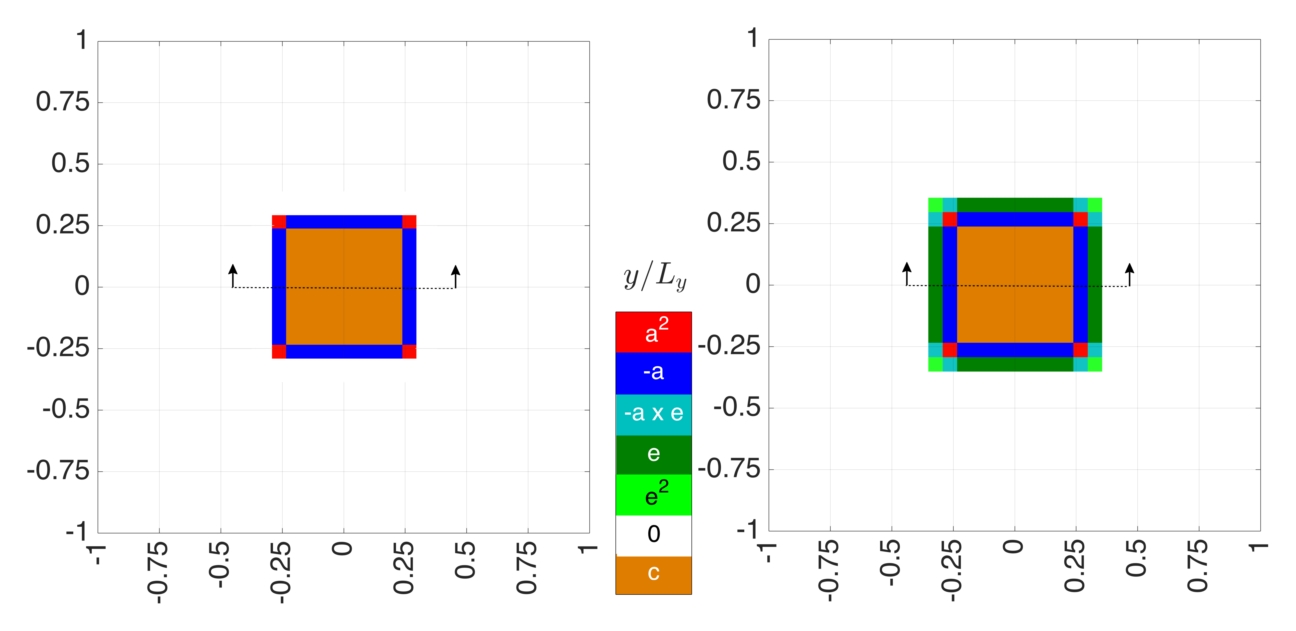}

	\caption{\footnotesize{Schematic of the SS kernels $\kMI(\bx)$ (left panels) and $\kMI(\bx)$ (right panels) we construct in 1D and 2D.}}	
\lb{fig:NewKernels}\end{figure}

It might be tempting to choose the free parameter $b$ to satisfy $\int dx~ x^4 \kMI(x) = 0$, however, it is straightforward to check that the solution is not realizable. Therefore, to construct a kernel $\kMII$ with a vanishing 4th moment in addition to the requirements on $\kMI$, 
we must infuse $\kMII$ with additional structure, shown in Fig. \ref{fig:NewKernels} as `arms' with a positive value $e$ and width $d = b$ on either side, thereby introducing one more parameter. In order for $\kMII$ to satisfy the two constraints, $\int dx~ x^2 \kMII(x) = 0$ and $\int dx~ x^4 \kMII(x) = 0$, we get 
\begin{eqnarray}
\left.\begin{aligned}
\frac{a}{c}&=&\frac{124\ b^3\  \ell^3+88\ b^2\  \ell^4+19\ b\  \ell^5+ \ell^6}{4\ b^2\left(192\ b^4+400\ b^3\ \ell+340\ b^2\ \ell^2+120\ b\ \ell^3+15\ \ell^4\right)}\lb{eq:MII_Aconstraint}\\
\frac{e}{c} &=&\frac{4\ b^3\ \ell^3+8\ b^2\ \ell^4+5\ b\ \ell^5+ \ell^6}{4\ b^2\left(192\ b^4+400\ b^3\ \ell+340\ b^2\ \ell^2+120\ b\ \ell^3+15\ \ell^4\right)} 
\end{aligned}\right\}(\kMII~\mbox{parameters} )
\end{eqnarray}
Similar to eq. (\ref{eq:MI_Aconstraint}), $b$ is a free parameter, which in this paper we take $b=\ell/8$ as in $\kMI$. This yields $a/c=568/257$ and $e/c= 200/257$ from eq. (\ref{eq:MII_Aconstraint}), and $c=\ell^{-1} 257/165$  from normalization. 

Note that the kernels become more complicated with increasing order. They also become increasingly spread over a wider range (longer stencils). While the width $b$ of the `limbs' (arms and legs in $\kMI$ and $\kMII$) is a free parameter that can be chosen by the user, an important practical consideration is that representing SS kernels on a grid requires at least 1 grid-cell of size $\Delta x$ for each of the limbs. Therefore, the smallest length-scale $\ell$ that can be probed by such a kernels is limited by $b\ge\Delta x$. On the other hand, if the limbs' width is made too large, the kernel becomes less localized in x-space and more expensive to use numerically.

The procedure described above can be followed to construct kernels of higher order. It is also straightforward to generalize any of these kernels to higher dimensions by defining it as a separable product;
for example in 2D, we define $G(x,y) \equiv G(x)G(y)$. If kernel $G(x)$ is of order $p$ in 1D, then $G(\bx)$ is of the same order $p$ in higher dimensions. Fig. \ref{fig:NewKernels} shows $\kMI$ and $\kMII$ in 2-dimensions.

\begin{figure}%[bhp]
\centering
		\includegraphics[width=0.85\textwidth,height=0.35\textwidth]{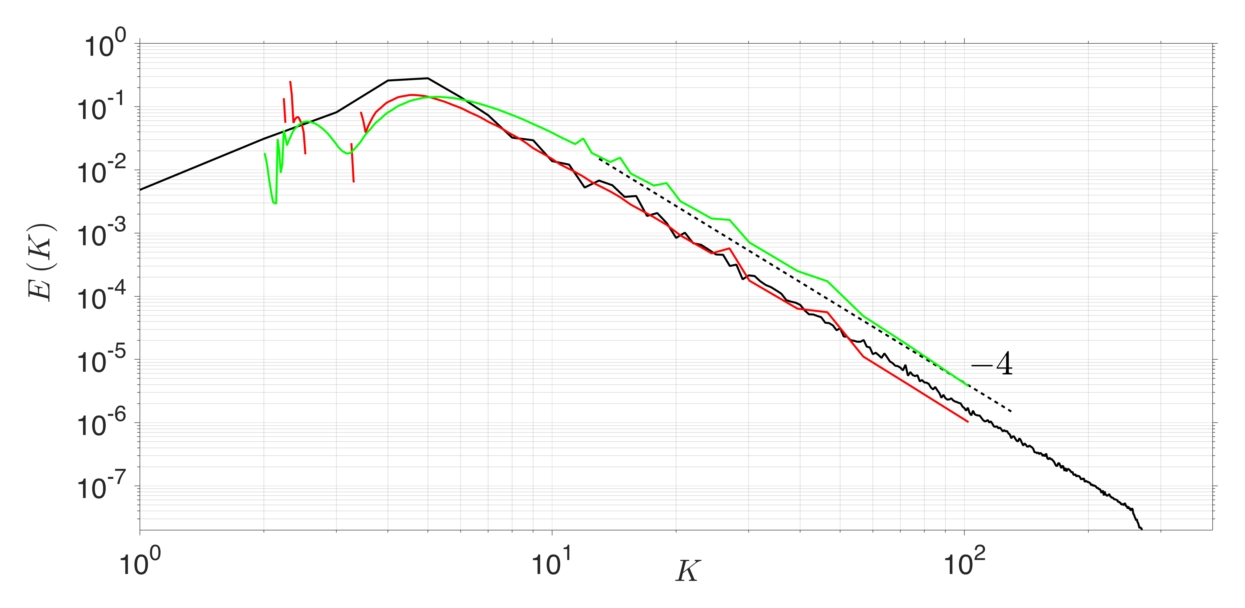}

	\caption{\footnotesize{ Energy spectrum for a doubly-periodic synthetic velocity field with a preset power-law slope of $K^{-4}$. Figure compares the Fourier spectrum $\left({\color{black}\textbf{---}}\right)$ to the filtering spectrum using kernels $\kMI \left({\color{green}\textbf{---}}\right)$ and  $\kMII \left({\color{red}\textbf{---}}\right)$, which we constructed above. Straight dashed black line with a $K^{-4}$ slope is for reference.}}	
\label{fig:NewKernelSpectrum}
\end{figure}

\begin{figure}%[bhp]
\centering
		\includegraphics[width=0.85\textwidth,height=0.35\textwidth]{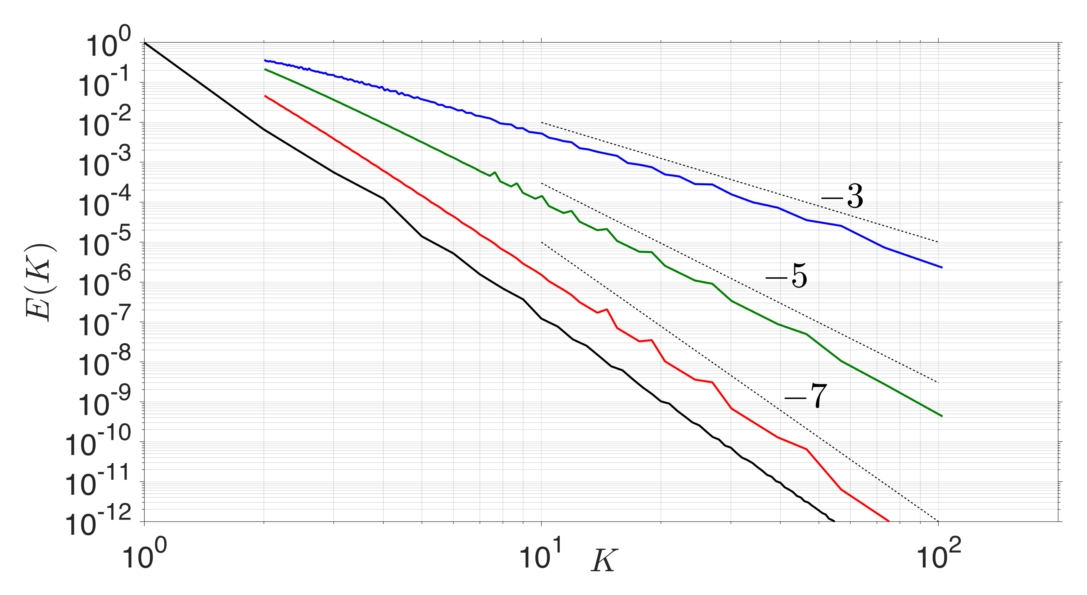}

	\caption{\footnotesize{Demonstrating the `locking' effect implied by our result (\ref{eq:FilterSpectrumScaling}) when calculating the filtering spectrum. We use a doubly-periodic synthetic velocity field with a preset spectral slope of $k^{-7}$.
We plot the filtering spectrum using Top-hat (${\color{blue}\textbf{---}}$), $\kMI$ (${\color[rgb]{0,.6,0}\textbf{---}}$), and $\kMII$ (${\color{red}\textbf{---}}$) kernels, along with the Fourier spectrum (${\color{black}\textbf{---}}$). We see that the filtering spectra using the three kernels (Top-hat, $\kMI$, $\kMII$) lock at $k^{-3}$, $k^{-5}$, and $k^{-7}$, respectively, consistent with eq. (\ref{eq:FilterSpectrumScaling}). 
	 }}	
\label{fig:NewKernelLimits}
\end{figure}

From eq. (\ref{eq:FilterSpectrumScaling}), the steepest slopes that can be measured by $\kMI$ and $\kMII$ are $\sim k^{-5}$ and $\sim k^{-7}$, respectively. Using a synthetic velocity field in a doubly periodic domain with a Fourier spectrum having a power-law scaling $\sim k^{-4}$, 
Fig. \ref{fig:NewKernelSpectrum} shows how the `filtering spectrum' using both kernels $\kMI$ and $\kMI$ can capture the spectral slope accurately whereas the Top-hat kernel locks at $k^{-3}$ in Fig. \ref{fig:GaussTopSpectra} due to its low order. $\kMII$ is slightly more accurate in capturing the wavenumber of spectrum's peak compared to $\kMI$.
Fig. \ref{fig:NewKernelLimits} shows how $\kMII$ can capture a spectral slope of $\sim k^{-7}$ correctly whereas $\kMI$ locks at $k^{-5}$, in agreement with relation (\ref{eq:FilterSpectrumScaling}). 

 \section{Application to Fluid Flows\lb{sec:ApplicationFlows}}
So far, we have tested the filtering spectrum on synthetic fields in periodic domains, having clear preset power-law scalings. 
In this section, we will apply the method to more realistic flows. We will show how a filtering spectrum can be used locally in x-space, over subregions of interest, and compare to Fourier-based techniques which alter the data to periodize it.

\subsection{2D decaying turbulence}
\begin{figure}[bhp]
\centering
	\includegraphics[width=.45\textwidth,height=0.45\textwidth]{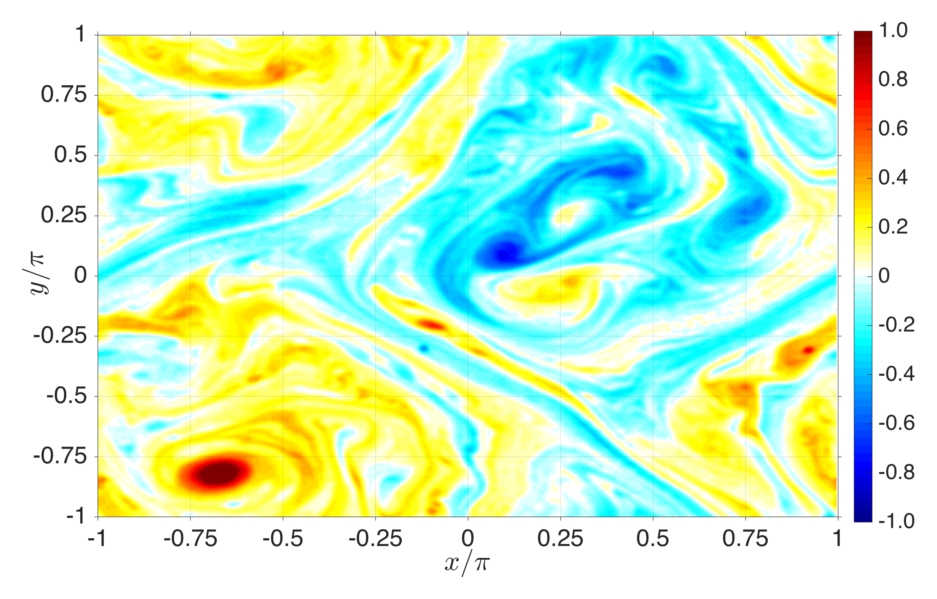}
		\includegraphics[width=.45\textwidth,height=0.45\textwidth]{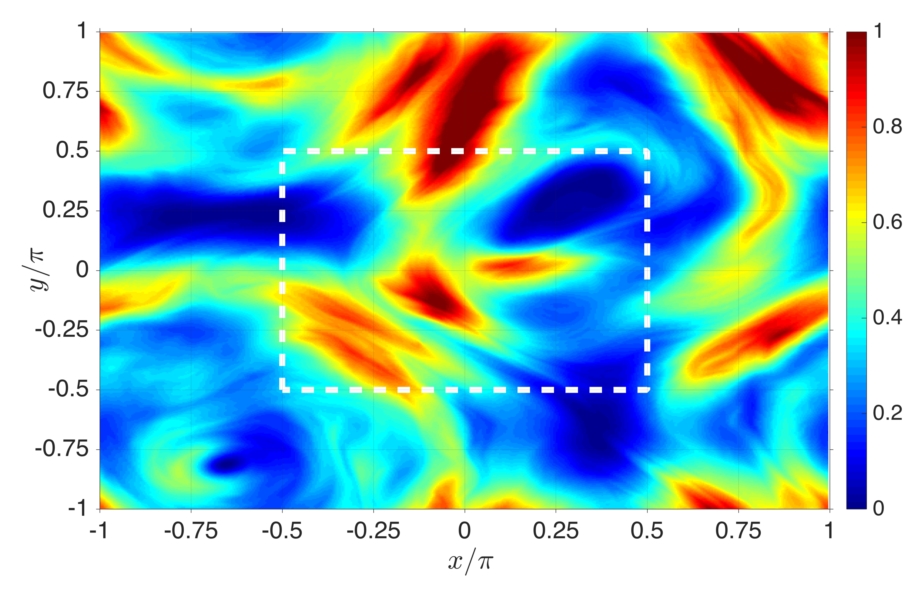}
	\includegraphics[width=0.45\textwidth,height=0.45\textwidth]{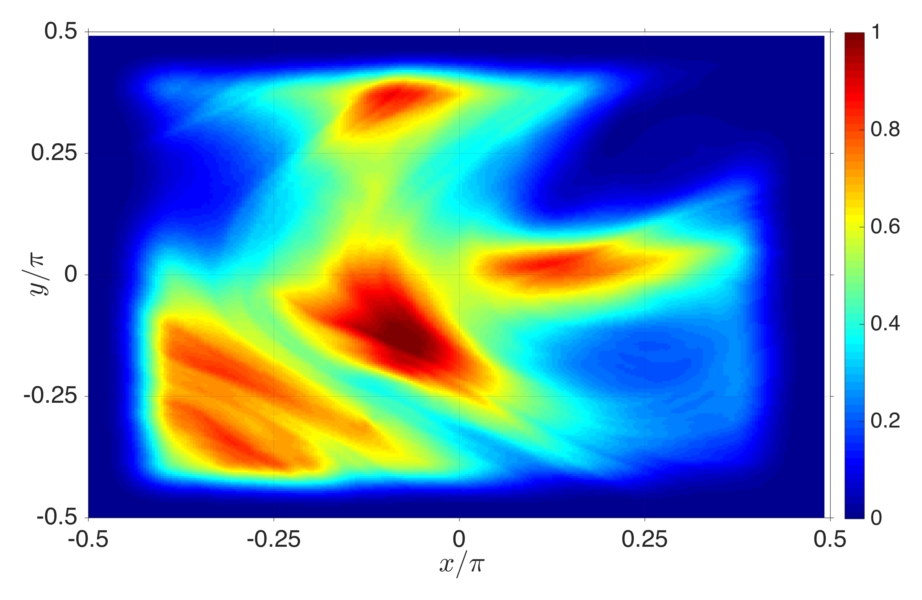}
		\includegraphics[width=0.45\textwidth,height=0.45\textwidth]{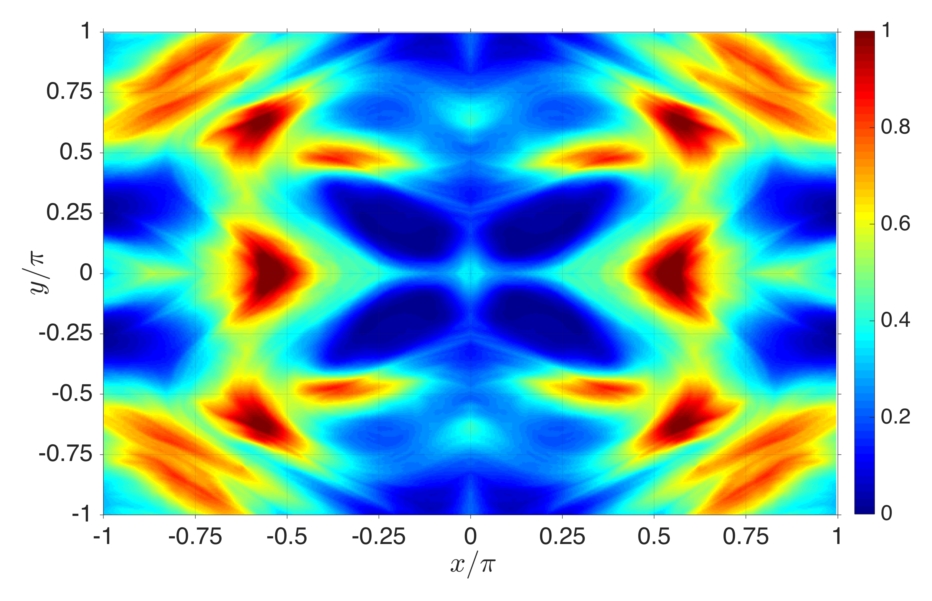}

	\caption{\footnotesize{Visualization of normalized vorticity (top-left) and kinetic energy (top-right) of the 2D periodic flow we analyze. The sub-domain is highlighted as a dashed white box. To calculate the Fourier spectrum over the sub-domain, the flow is periodized by tapering (lower-left) or mirroring (lower-right).
}}
\label{fig:2Dturb}
\end{figure}

\begin{figure}[bhp]
\centering
				\includegraphics[width=.8\textwidth,height=0.3\textheight]{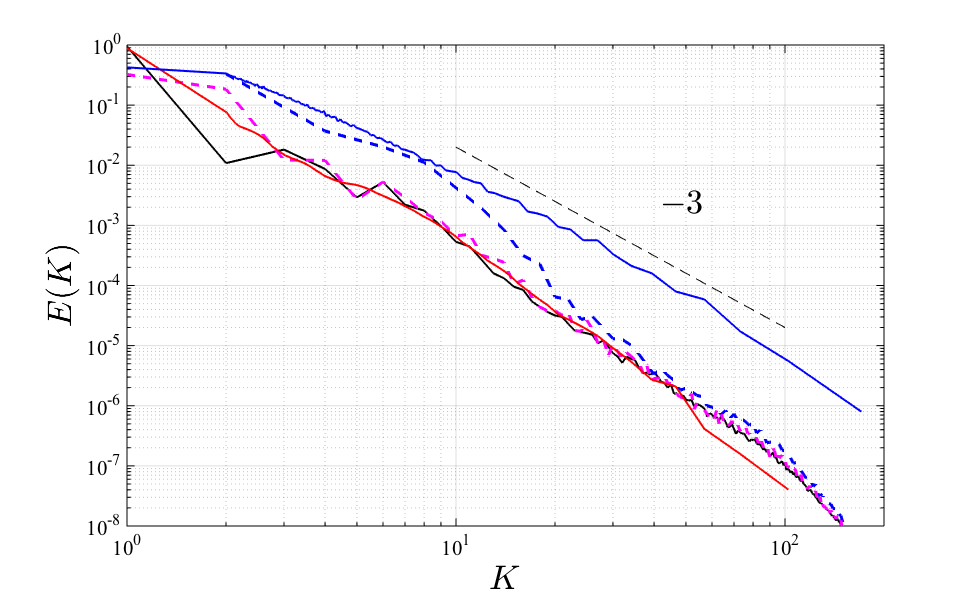}

	\caption{Figure of spectra from a doubly-periodic 2D flow, normalized by the spatial average of energy over the entire domain. It shows the Fourier spectrum over the entire domain $\left(  {\color{black}\textbf{---}}   \right)$ decaying slightly steeper than $k^{-3}$. It also shows the Fourier spectrum over the sub-domain highlighted in Fig. \ref{fig:2Dturb}, calculated by the tapering $\left( {\color{blue}\textbf{- -}} \right)$ and mirroring $\left( {\color{magenta}\textbf{- -}} \right)$ methods. We compare to the filtering spectra using a Top-hat kernel $\left(  {\color{blue}\textbf{---}}   \right)$ and the $\kMII$ kernel $ \left( {\color{red}\textbf{---}} \right)$. 
Note that tapering, while it reduces the total energy in the sub-domain, redistributes some energy from the largest scales to smaller scales, creating a significant bump over $k\in[2,20]$. The filtering spectrum with a Top-hat kernel `locks' at a $k^{-3}$ scaling as expected for this flow.  Both the filtering spectrum with the $\kMII$ kernel, and the Fourier spectrum using mirroring yield spectra that follow closely the Fourier spectrum over the entire domain.}	
\label{fig:2Dspectrum}
\end{figure}

We analyze the velocity field $\bu$ generated 
from a direct numerical simulation of the incompressible Navier Stokes equation,
\be  \partial_{t} \bu +  (\bu\cdot\grad)\bu = -\grad p + \nu_3 \nabla^6 \bu, \hspace{1cm}\grad\cdot\bu=0,
\lb{NS-eq} \ee
which is solved pseudo-spectrally in a  in a doubly periodic domain $[-L/2,L/2)^2 = [-\pi,\pi)^2$
on a $512^2$ grid. Here, 
$p$ is the pressure, $\nu_3=5\times10^{-14}$ is the hyperviscosity. 
The flow is randomly initialized
such that the initial energy and enstrophy are at the large scales and we resolve the forward 
enstrophy cascade range where the energy spectrum is expected to scale as $E(k)\sim k^{-3}$ \cite{Batchelor69,Kraichnan67,KraichnanMontgomery80,BoffettaEcke12}.
The energy spectrum we measure in Fig. \ref{fig:2Dspectrum} is consistent with that scaling, albeit slightly steeper.
We analyze a snapshot of the flow visualized in Fig. \ref{fig:2Dturb}. 

In order to show how the filtering spectrum performs over small regions in the flow which may be of interest in an application, 
we choose a sub-domain box of size $L/2\times L/2$ shown in Fig. \ref{fig:2Dturb}. Since the flow over the sub-domain is not periodic,
measuring the Fourier spectrum requires either tapering the flow near the edges or mirroring (reflecting) the sub-domain to periodize the data. Fig. \ref{fig:2Dturb} visualizes the flow resulting from these two methods. A Fourier spectrum can then be measured from each of the two periodized velocity fields, as shown in Fig. \ref{fig:2Dspectrum}. Note that with the tapering method, the largest length-scale is $L/2$, which corresponds to a smallest wavenumber of $k=2$ in Fig. \ref{fig:2Dspectrum}. The mirroring method, in this case, yields a ``super-box'' of the same size as the original domain (this is not generally true). We compare these with the filtering spectrum using the Top-hat and $\kMII$ kernels, which can be applied over the sub-domain without having to periodize the data. 

\clearpage

Since the flow is statistically homogenous and the sub-domain covers a significant portion of the domain, we expect  the local spectrum to be similar to the global Fourier spectrum. Both the filtering spectrum using $\kMII$ and the Fourier spectrum evaluated over the ``super-box'' yield spectra fairly similar to the global Fourier spectrum, with the latter performing slightly better, especially at small scales. However, it can be seen from Fig. \ref{fig:2Dturb} that mirroring the flow introduces an artificial periodic pattern at the largest scales, which appears in the ``super-box'' Fourier spectrum at $k=2$ in Fig. \ref{fig:2Dspectrum}. Below, we shall discuss an example where the spurious artifacts from mirroring dominate the Fourier spectrum.
In comparison, the windowed Fourier spectrum obtained by tapering does not perform as well with significant deviations at the large scales and in the power-law scaling. This is due to the artificial gradients introduced by tapering which can alter the spectral content. 

\begin{figure}%[bhp]
\centering
	\includegraphics[width=.45\textwidth,height=0.45\textwidth]{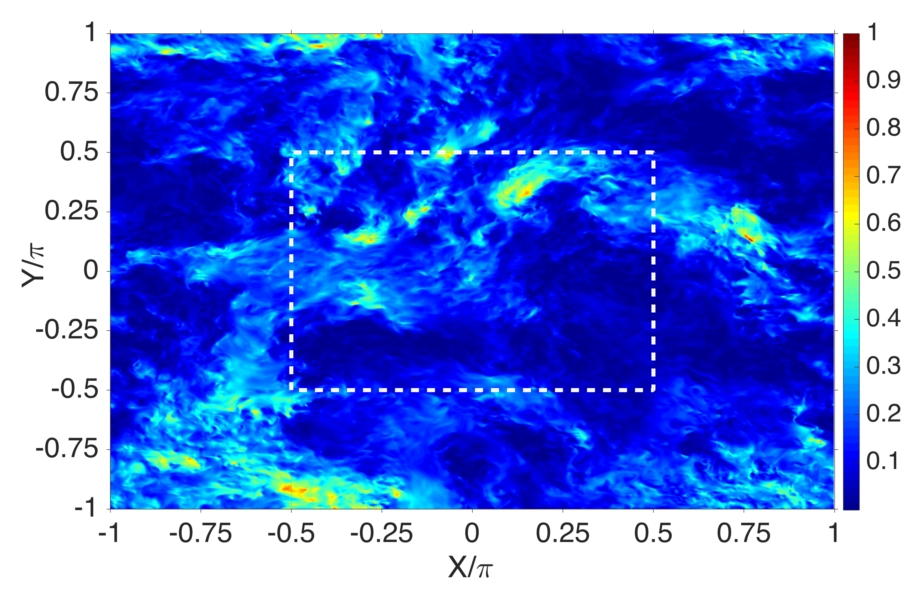}
		\includegraphics[width=.54\textwidth,height=0.45\textwidth]{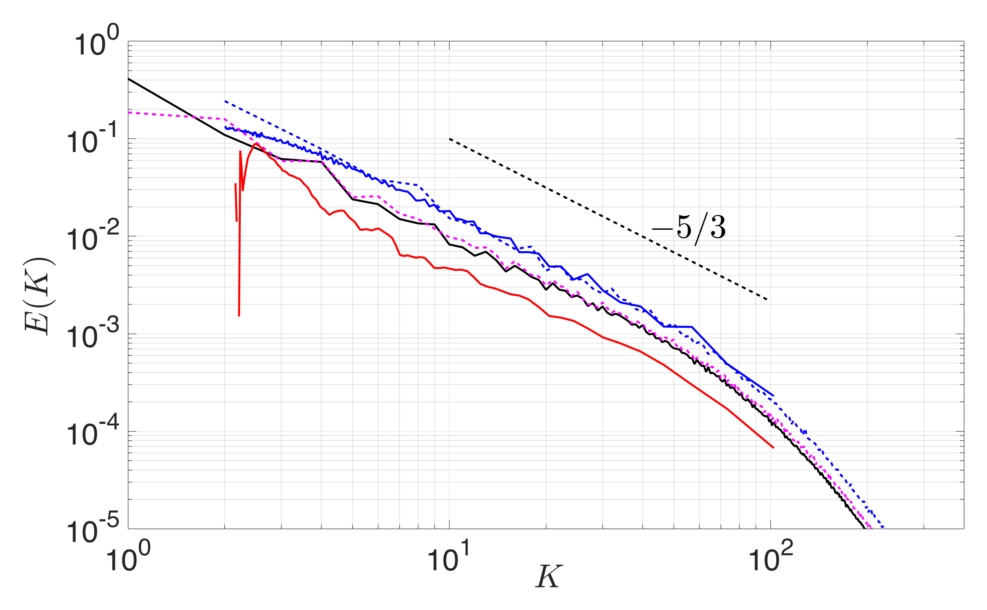}
	\caption{\footnotesize{
	A sample 2D slice of the normalized kinetic energy of a 3D flow from the JHTDB (left panel). The dashed white box bounds the sub-domain of interest. Right panel shows the Fourier spectrum over the entire domain $\left(  {\color{black}\textbf{---}}   \right)$. It also shows the Fourier spectrum over the sub-domain periodized by the tapering $\left( {\color{blue}\textbf{- -}} \right)$ and mirroring $\left( {\color{magenta}\textbf{- -}} \right)$ methods. We compare to the filtering spectra using a Top-hat kernel $\left(  {\color{blue}\textbf{---}}   \right)$ and the $\kMII$ kernel $ \left( {\color{red}\textbf{---}} \right)$. 	All spectra are normalized by the spatial average of energy over the entire domain.}}	
\label{fig:JHUturb}
\end{figure}

Fig. \ref{fig:JHUturb} shows a similar analysis on $1{,}024^2$ slices of 3D forced isotropic turbulence obtained from the JHU turbulence database \cite{Lietal08}. Here, we expect a $k^{-5/3}$ scaling of the spectrum associated with a forward energy cascade. We calculate the global Fourier spectrum, along with Fourier spectra obtained from the sub-domain shown using the two periodization methods, mirroring and tapering. We also calculate the filtering spectrum over the sub-domain and observe a putative power-law scaling very similar to that of the global Fourier spectrum. However, the Fourier spectrum obtained with mirroring seems to match most closely the global Fourier spectrum. The Fourier spectrum obtained with tapering suffers from problems similar to those we discussed in Fig. \ref{fig:2Dspectrum}.

\begin{figure}%[bhp]
\centering
	\includegraphics[width=.43\textwidth,height=0.43\textwidth]{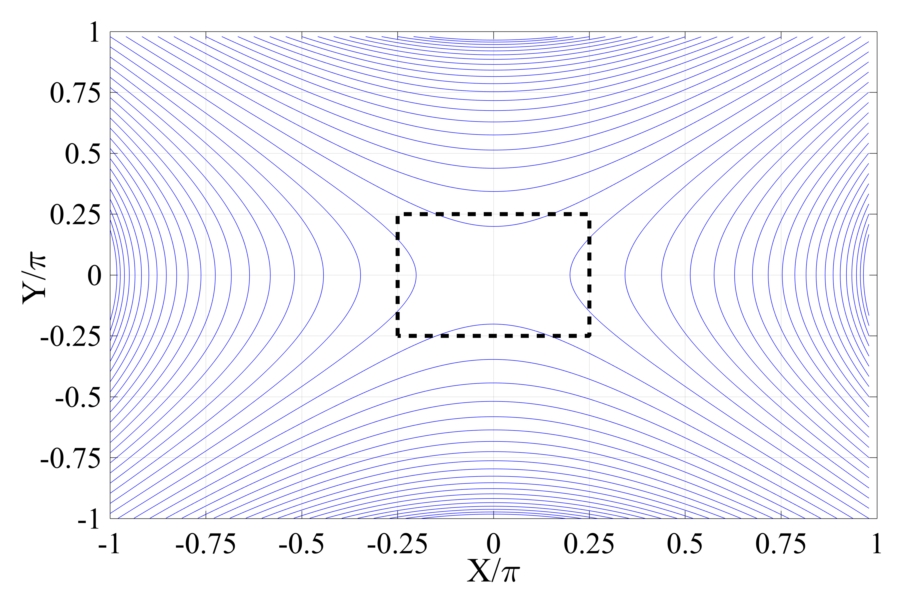}
	\includegraphics[width=0.55\textwidth,height=0.42\textwidth]{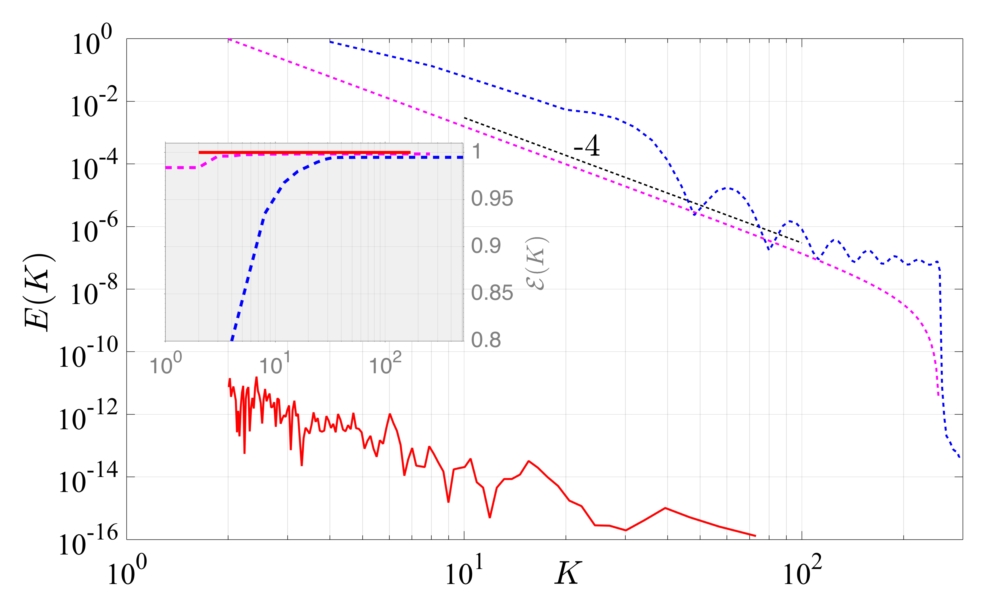}
	\includegraphics[width=1\textwidth,height=0.35\textwidth]{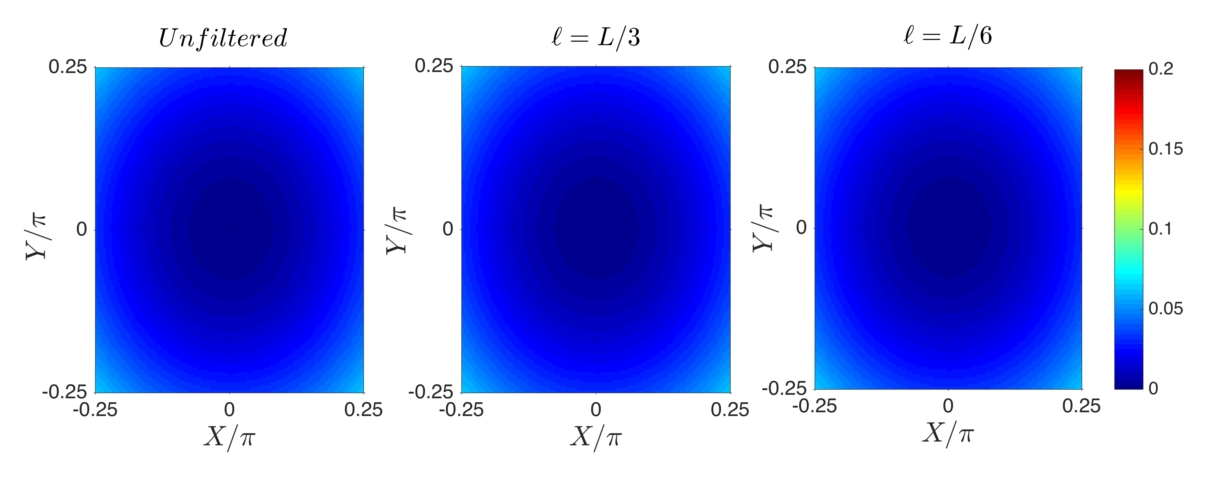}
	\caption{\footnotesize{ 
	Streamlines of a large scale strain defined by $\bu =(y,x)$ (top-left panel). We calculate the spectral content in the sub-domain $[-\pi/4,\pi/4]^2$ (black dashed box). Top-right panel shows the filtering spectrum over the sub-domain using $ \kMII$ $\left(  {\color{red}\textbf{---}} \right)$, and the Fourier spectrum by the mirroring $\left(  {\color{magenta}\textbf{- -}} \right)$ and tapering $\left( {\color{blue}\textbf{- -}} \right)$ methods. Inset plots the cumulative spectrum, $\mE(k)$. Here, $k=L/\ell$, with $L=2\pi$. Bottom three panels show the kinetic energy of the unfiltered flow, $|\bu|^2/2$ (left), and that of the filtered flow $|\OL\bu_\ell|^2/2$ at $\ell=L/3$ (middle) and $\ell=L/6$ (right). All three bottom panels look almost identical, indicating  the absence of small scales, consistent with the flow being a large-scale smooth coherent strain. This is revealed in the filtering spectrum, which shows almost zero spectral content at all small scales probed. In contrast, the Fourier spectrum is almost 12 orders of magnitude larger, exhibiting a power-law over a continuum of scales due to the spurious patterns that arise from mirroring. Both the Fourier spectrum and the filtering spectrum measure the same total energy as evidenced from $\mE(k)$ at large $k$ in the inset of top-right panel.}}	
\label{fig:Strain}
\end{figure}

To illustrate a situation in which the Fourier spectrum obtained by the mirroring method can yield misleading results, we consider a large-scale strain field 
\be
u_x = y, \hspace{1cm} u_y=x,
\ee 
shown in Fig \ref{fig:Strain}. This flow is not periodic and is \emph{large-scale} in the sense that its derivative, the strain field, is a constant. Equivalently, the velocity components $u_x(y)$ and $u_y(x)$ are linear functions with no small-scale variation. Therefore, we should expect a spectrum to reveal zero spectral content at small scales. In fact, Fig \ref{fig:Strain} shows that filtering the smooth flow at two different scales yields absolutely no change to the unfiltered flow, indicating the absence of small scales. This is consistent with the filtering spectrum  (see Fig. \ref{fig:Strain}) over the region $[-\pi/4,\pi/4]^2$, which shows almost zero energy at small-scales. In contrast, the Fourier spectrum obtained by mirroring exhibits a $k^{-4}$ power-law over a continuum of scales, which is an artifact of the periodic patterns that arises from mirroring. This example highlights that for flows in which there is a strong coherent smooth flow with weak turbulence, or a significant separation between the large scales and the turbulence such as in Rapid Distortion Theory, measuring the spectrum with Fourier methods can yield misleading results.

 \clearpage

 \section{Conclusion\lb{sec:Conclusion}}
We have shown that the power-law spectrum in a turbulent flow can be extracted by a relatively simple procedure of low-pass filtering in x-space. We have also shown that for a flow with a certain level of regularity or smoothness (as quantified by the steepness of its spectrum), the filtering kernel must have a sufficient number of vanishing moments in order for the filtering spectrum to be meaningful.
The smoother is the flow (steeper is the spectrum), the more complicated the filtering kernel has to be. Since most spectra encountered in turbulence are not smooth, having a slope shallower than $k^{-3}$, a simple averaging of adjacent grid-points, equivalent to filtering with a Top-hat kernel, can uncover the power-law spectrum of the flow. This can be done with a few lines of code, which we believe is a main appeal of the method. On the other hand, if the filtering spectrum using a Top-hat  yields a power law scaling of $k^{-3}$, then slightly more complicated SS kernels constructed above can be used to extract the spectrum. These SS kernels are also quite straightforward to implement on a numerical grid.

An important advantage of the method presented here over the wavelet spectrum is that it can be used to calculate a generalized ``spectrum'' of any quantity that is non-quadratic. The wavelet spectrum relies on Plancherel's relation to conserve energy and, therefore, requires treating quantities as quadratic even when they are not. In contrast, the filtering spectrum conserves energy due to the Fundamental Theorem of Calculus. In forthcoming work \cite{Zhaoetal_inprep}, we will show how it can be used to extract a generalized ``spectrum'' of energy in compressible or variable density flows, when such energy is cubic as emphasized in \cite{Aluie11,EyinkDrivas18,ZhaoAluie18}.

 \section*{Acknowledgement}
The authors are grateful to M. Hecht and G. Vallis for discussions that motivated this research. We also thank two anonymous referees for their valuable comments which helped improve the paper.
This work was supported by NASA grant 80NSSC18K0772 and the DOE Office of Fusion Energy Sciences grant DE-SC0014318. HA was also partially supported by the DOE National Nuclear Security Administration under award DE-NA0001944, the University of Rochester, and the New York State Energy Research and Development Authority. HA also thanks KITP for its hospitality with the support of NSF grant PHY17-48958. This research used resources of the National Energy Research Scientific Computing Center, a DOE Office of Science User Facility supported by the Office of Science of the U.S. Department of Energy under Contract No. DE-AC02-05CH11231. 
Part of our study used data from the Johns Hopkins Turbulence Database at \url{http://turbulence.pha.jhu.edu}. A sample code for using the method described here can be found on our group's website, \url{http://www.complexflowgroup.com/}

 \renewcommand{\theequation}{A-\arabic{equation}}
\setcounter{equation}{0}  % reset counter 
\setcounter{Prop}{0}  % reset counter 
\section*{APPENDIX}  % use *-form to suppress numbering
\subsection{Filtering Spectrum Scaling}
We first discuss the scaling in eq. (\ref{eq:FilterSpectrumScaling}).
Assume that $E(k)= \const k^{-\alpha}$ over $k_a<k<\infty$, then we have
\begin{eqnarray}
\OL{E}(k_\ell)&=&\int_0^{\infty} dk~ \frac{d}{dk_\ell}  \left|\wh{G}\left(\frac{k}{k_\ell}\right)\right|^2 E(k) \nonumber\\
&=&\underbrace{ \int_0^{k_a} dk~ \frac{d}{dk_\ell}  \left|\wh{G}\left(\frac{k}{k_\ell}\right)\right|^2 E(k)}_{\mbox{\footnotesize{term 1}}} + \underbrace{\const \int_{k_a}^{\infty} dk~ \frac{d}{dk_\ell}  \left|\wh{G}\left(\frac{k}{k_\ell}\right)\right|^2 k^{-\alpha} }_{\mbox{\footnotesize{term 2}}}
\end{eqnarray}
The small wavenumber contributions to the filtering spectrum $\OL{E}(k_\ell)$ at any $k_\ell$ are captured by `term 1', which will be shown to scale as $\sim k_\ell^{-(p+2)}$, whereas the high wavenumber contributions in `term 2' scale as $\sim k_\ell^{-\alpha}$ and reflect the scaling of the Fourier spectrum.
Therefore, if the Fourier spectrum decays faster than $k^{-(p+2)}$, the small-wavenumber contributions in `term 1' dominate the scaling of the `filtering spectrum' at large $k_\ell$, whereas if $\alpha < p+2$, then the `filtering spectrum' has the same power-law slope as the Fourier spectrum.

Term 2 can be rewritten with a change of variable $s=k/k_\ell$:
\begin{eqnarray}
 \int_{k_a}^{\infty} dk~ \frac{d}{dk_\ell}  \left|\wh{G}\left(\frac{k}{k_\ell}\right)\right|^2 k^{-\alpha} &=& k_\ell^{-\alpha} \int_{k_a/k_\ell}^{\infty} ds \frac{d}{ds} \left|\wh{G}\left(s\right)\right|^2 s^{1-\alpha}.
\end{eqnarray}
Under mild smoothness and decay conditions on $\wh{G}\left(s\right)$, the integral on the right hand side converges to a constant and `term 2' scales as $k_\ell^{-\alpha}$.

To analyze the scaling of `term 1,' we use the Taylor series expansion of $\wh{G}(s)$ from eq. (\ref{eq:TaylorExpandKernel}) with $s=k/k_\ell$ such that
\begin{eqnarray}
\frac{d}{dk_\ell}  \left|\wh{G}\left(\frac{k}{k_\ell}\right)\right|^2 
&=& -2 \left[1+\left(\frac{k}{k_\ell}\right)^{p+1} \phi\left(s\right)\right] 
\left[ (p+1) \frac{k^{p+1}}{k_\ell^{p+2}}\,\phi\left(s\right) +  \frac{k^{p+2}}{k_\ell^{p+3}}\,\phi^{(1)}\left(s\right)\right]  
\nonumber\\
&\sim& \underbrace{k_\ell^{-(p+2)} k^{p+1} \phi\left(s\right)}_{\mbox{term a}} + \underbrace{k_\ell^{-(p+3)} k^{p+2}\,\phi^{(1)}\left(s\right)}_{\mbox{term c}}
+ \underbrace{k_\ell^{-(2p+3)} k^{2p+2}\left|\phi\left(s\right)\right|^2}_{\mbox{term b}} \nonumber\\
 && + \underbrace{k_\ell^{-(2p+4)} k^{2p+3}\,\phi\left(s\right)\,\phi^{(1)}\left(s\right)}_{\mbox{term d}}
\end{eqnarray}

\noindent
Consider the first of these small wavenumber contributions for large $k_\ell$:
\begin{eqnarray}
\mbox{term 1a} &=&  \int^{k_a}_{0} dk~ \left[k_\ell^{-(p+2)} k^{p+1} \phi\left(\frac{k}{k_\ell}\right)\right] E(k) \nonumber \\
&\approx& k_\ell^{-(p+2)}  \int^{k_a}_{0} dk\, \const k^{p+1} E(k) \nonumber \\
&\sim& k_\ell^{-(p+2)} \nonumber 
\end{eqnarray}
where  we used eq.(\ref{eq:PhiProperty0}) in the second step, that $\phi(k/k_\ell)\approx \const$ when $k_\ell\to\infty$.

Similar analysis on the other terms yields that the other terms 1b, 1c, and 1d are subdominant to $k_\ell^{-(p+2)}$ for large $k_\ell$. Therefore, we have that
\begin{eqnarray}
\OL{E}(k_\ell)&=&\underbrace{ \int_0^{k_a} dk~ \frac{d}{dk_\ell}  \left|\wh{G}\left(\frac{k}{k_\ell}\right)\right|^2 E(k)}_{\sim k_\ell^{-(p+2)}} + \underbrace{\const \int_{k_a}^{\infty} dk~ \frac{d}{dk_\ell}  \left|\wh{G}\left(\frac{k}{k_\ell}\right)\right|^2 k^{-\alpha} }_{\sim k_\ell^{-\alpha}}
\end{eqnarray}

\subsection{Positive Definiteness}
We now discuss the sign of the kernel $\Gamma(\br) = G_\ell^{-1}*G_\Delta$ used in eq. (\ref{eq:SoftDeconv}), with $\ell\le\Delta$. We assume that the filtering kernel $G(\br)$ is a concave function (and therefore $G\ge0$). The soft deconvolution can be written as an expansion \cite{Sagaut06}:
\begin{eqnarray} 
\Gamma(x)=G_\ell^{-1}*G_\Delta(x) &=& \sum_{k=0}^{\infty} (I-G_\ell)^k*G_\Delta \approx  G_\Delta +  (G_\Delta-G_\ell*G_\Delta),
\lb{eq:App:GammaExpand}\end{eqnarray}
where $I$ is the identity operator and we assume that the series converges sufficiently fast to justify truncating the expansion. We then have
\begin{eqnarray} 
G_\Delta-G_\ell*G_\Delta &=& -\int dr\,G_\ell(r) [G_\Delta(x+r)-G_\Delta(x)] \nonumber\\
&\approx&  -  \frac{\ell^2}{\Delta^2}(\nabla^2G)_\Delta  \int dr\, r^2 G(r),
\end{eqnarray}
where we Taylor expanded the even kernel $G_\Delta(x+r)$ near $x$ and $(\nabla^2G)_\Delta = \frac{1}{\Delta^d}\frac{\partial^2G(x/\Delta)}{\partial(x/\Delta)^2}$ in $d$-dimensions. Since the kernel is concave, then $-(\nabla^2G)_\Delta\ge0$ and we have
\be \Gamma(x) = G_\ell^{-1}*G_\Delta(x) \approx  G_\Delta +  (G_\Delta-G_\ell*G_\Delta) \ge0. 
\ee
Note that in the limit $\ell\ll\Delta$, $(G_\Delta-G_\ell*G_\Delta)$ in eq. (\ref{eq:App:GammaExpand}) is proportional to $(\ell/\Delta)^2$ and, therefore,
$\Gamma \approx G_\Delta\ge0$. The reader should realize that the analysis just presented is not a rigorous proof but an argument which relies on significant approximations. We speculate that it may be possible to show positivity of $\Gamma$ using a more careful analysis which does 
not, for example, require $G(r)$ to be concave but only that it is positive.

\clearpage
\newpage
%\end{list}

\bibliographystyle{unsrt}
 \bibliography{FilteringSpectrum}

\end{document}